\documentstyle[12pt,psfig]{article}
\font\bbexii=cmmib10 scaled\magstep1
\font\bbex=cmmib10
\font\bbevii=cmmib7

\newfam\bbefam
\textfont\bbefam=\bbexii
\scriptfont\bbefam=\bbex
\scriptscriptfont\bbefam=\bbevii
\mathchardef\BigEchar="710F

\begin{document}
\def\lag{\langle}
\def\rag{\rangle}
\begin{titlepage}
\vspace{2.5cm}
\baselineskip 24pt
\begin{center}
\vspace{1.5cm}
\vspace{1.5cm}
\large\bf {Signal Confidence Limits from a Neural Network Data
Analysis$^1$}\\
\vspace{1.5cm}
\large{
Bernd A. Berg$^{2,3,4,5}$ and J\"urgen Riedler$^{2,4,6,7}$
}
\end{center}
\vspace{2cm}
\begin{center}
{\bf Abstract}
\end{center}
This paper deals with a situation of some importance for
the analysis of experimental data via Neural Network (NN) or similar
devices: Let $N$ data be given, such that \hbox{$N=N_s+N_b$}, where $N_s$ is 
the number of signals, $N_b$ the number of background events, both 
unknown. Assume that a NN has been trained, such that it 
will tag signals with efficiency $F_s$, $(0<F_s<1)$ and background 
data with $F_b$, $(0<F_b<1)$. Applying the NN yields $N^Y$ tagged 
events. We demonstrate that the knowledge of $N^Y$ is sufficient to
calculate confidence bounds for the signal likelihood, which have the
same statistical interpretation as the Clopper-Pearson bounds for the
well-studied case of direct signal observation.

Subsequently, we discuss rigorous bounds for the {\it a-posteriori}
distribution function
of the signal probability, as well as for the (closely related) 
likelihood that there are $N_s$ signals in the data. We compare them
with results obtained by starting off with a maximum entropy type 
assumption for the {\it a-priori} likelihood that there are $N_s$ 
signals in the data and applying the Bayesian theorem. Difficulties 
are encountered with the latter method.
\vfill

\footnotetext[1]{{This research was partially funded by the
Department of Energy under contract DE-FG05-87ER40319\hfill
{\hskip 10pt and} by the Austrian Ministry of Science.}}
\footnotetext[2]{{Department of Physics, The Florida State University,
                      Tallahassee, FL~32306, USA. }}
\footnotetext[3]{{Supercomputer Computations Research Institute,
                      Tallahassee, FL~32306, USA.}}
\footnotetext[4]{{Zentrum f\"ur Interdisziplin\"are Forschung (ZIF),
 Wellenberg 1, D-33615 Bielefeld, Germany.}}
\footnotetext[5]{{E-mail: berg@hep.fsu.edu}}
\footnotetext[6]{{Institut f\"ur Kernphysik, Technische Universit\"at Wien,
 A-1040 Vienna, Austria.}}
\footnotetext[7]{{E-mail: juri@kph.tuwien.ac.at}}
\end{titlepage}

\baselineskip 24pt

\section{Introduction}

Let us assume that $N_s$ signals are observed in $N$ data
$$ N = N_s + N_b\, , $$
where $N_b$ is the number of background events. We denote the {\it a-priori}
unknown signal likelihood by $p$.  Relying on the binomial distribution, 
Clopper and Pearson~\cite{ClPe} derived a method, which allows to calculate 
rigorous confidence bounds on $p$, given $N_s$ and $N$.
Now, in modern physics, in particular high energy physics experiments, it 
happens quite often that signal and background events belong to overlapping 
probability densities in a multi-dimensional parameter space. In such
situations signals can only be identified in a statistical sense.
Typically, some method may allow to tag signals and background events with 
different efficiencies: $F_s$ for the signals $(0<F_s<1)$ and $F_b$ for
the background events $(0<F_b<1)$. Instead of observing $N_s$ signals we 
only get
$$ N^Y ~~{\rm tagged\ data.}$$
The question is, what confidence limits on the signal likelihood are then 
implied? We proof, and illustrate in some detail, that the Clopper-Pearson 
method can be generalized accordingly.

In particular, we have high energy physics experimental data in mind, where 
tagging may be provided by traditional cuts or by applying some
NN~\cite{Loe94,BeBl93,StYe93,RiLi91,Be96} technique. To give an example,
figure~5 of Ref.\cite{Be96} depicts values for neural network efficiencies, 
$F_s(Y)$ and $F_b(Y)$, which one may expect to occur for identifying
$t\bar{t}$-events in the All-Jets channel \cite{D0}.
Running the network on all $N$ data assigns to each event a value
of the network function $Y_n$, $n=1,...,N$. For a fixed choice of $Y$, 
the network returns $N^Y$ events with $Y_n\le Y$. An additional problem 
in real applications may be that the efficiencies $F_s$ and $F_b$ are not 
exact either. However, as outlined in the conclusions, we think that this
difficulty may be overcome by the bootstrap approach~\cite{Ef87}.

In the next section we explain and generalize the Clopper-Pearson approach.
A number of illustrations focus on the small number
of $N=10$ data, because then  the statistical meaning of the confidence bounds
becomes most transparent. In section~3 we deal with the limit of a 
small number of signals hidden in a large data set. 
Two instructive sets of network efficiencies are chosen, to 
demonstrate how the general equations are expected to work in practice.

Section~4 considers {\it a-posteriori} distribution functions. (i) For the 
signal probability
$$ F(p)\ =\ \int_0^p \rho (p')\, dp'\, $$
where $\rho (p)$ is the probability density of $p$. (ii) For the 
likelihood that there are $N_s$ signals in the data set
$$ F(N_s)\ =\ \sum_{k=0}^{N_s} P(k)\, ,$$
where $P(N_s)$ is the probability that there are $N_s$ signals in the 
data data. Rigorous lower and upper bounds are provided. For the examples
of section~3 those bounds are close together, such that useful
approximations of the true {\it a-posteriori} distribution functions
result. In section~5 these results are compared with constructing
the $F(N_s)$ distribution function and its $P(N_s)$ probability density
with the Bayesian method under the maximum entropy assumption that each $N_s$
is, {\it a-priori}, equally likely. One of the obtained results, and
hence its {\it a-priori} assumption, is in violation to an exact 
bound. Conclusions follow in the final section~6.
\hfill\break

\section{From Neural Network Output to Confidence Limits}

Let $p$ be the (unknown) exact likelihood that a data point is a signal. 
The probability to observe $N_s$ signals within $N$ measurements is given
by the binomial probability density
\begin{equation} \label{binomial}
b(N_s|N,p) = \left(\!\!\begin{array}{c}
                            N \\
                            N_s
                     \end{array}\!\!
               \right) p^{N_s}\, q^{N-N_s}\, ,\ q=1-p\, .
\end{equation}
We are faced with the inverse problem: if $N_s$ signals are observed, 
what is the confidence to rule out certain $p$? 
Assume that probabilities $p^c_-$ and $p^c_+$ are given. 
Clopper and Pearson~\cite{ClPe} define corresponding lower $p_-$
and upper $p_+$ bounds as solutions of the equations
\begin{equation} \label{CPbounds}
  p^c_-=\sum_{k=N_S}^{N} b(k|N,p_-)\ ~~{\rm and}\ 
~~p^c_+=\sum_{k=0}^{N_s} b(k|N,p_+) 
\end{equation}
with the additional convention $p_-=0$ for $N_s=0$ and $p_+=1$ for
$N_s=N$. Figure~1 illustrates, how $p_-$ and $p_+$ are obtained as 
parameters of the binomial distributions which yield the areas $p^c_-$ 
and $p^c_+$ as indicated. For this figure we have chosen $N=26k$ and
$N_s=130$, in the ballpark of values which will interest us in the next
section. Here and in the following binomial coefficients have been
calculated relying on Fortran routines of~\cite{NumR}. 

\begin{figure}[b] \label{bicomp}
\centerline{\psfig{figure=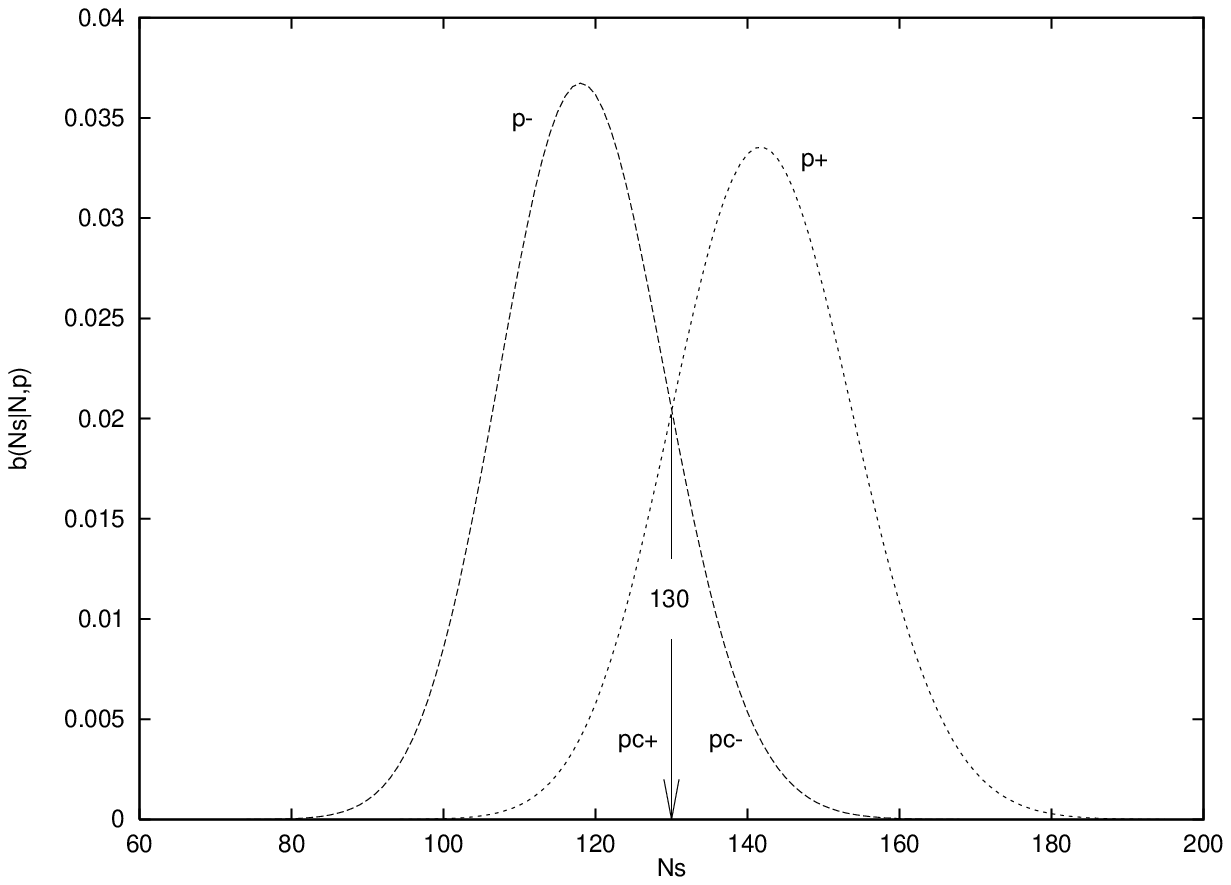}}
\caption{Binomial probability densities corresponding to solving 
equations~(\ref{CPbounds}) for $p_+$ and $p_-$ with $p^c_-=p^c_+=0.159$, 
$N=26k$ and $N_s=130$.}
\end{figure}

\begin{figure}[b] \label{CPprob}
\centerline{\psfig{figure=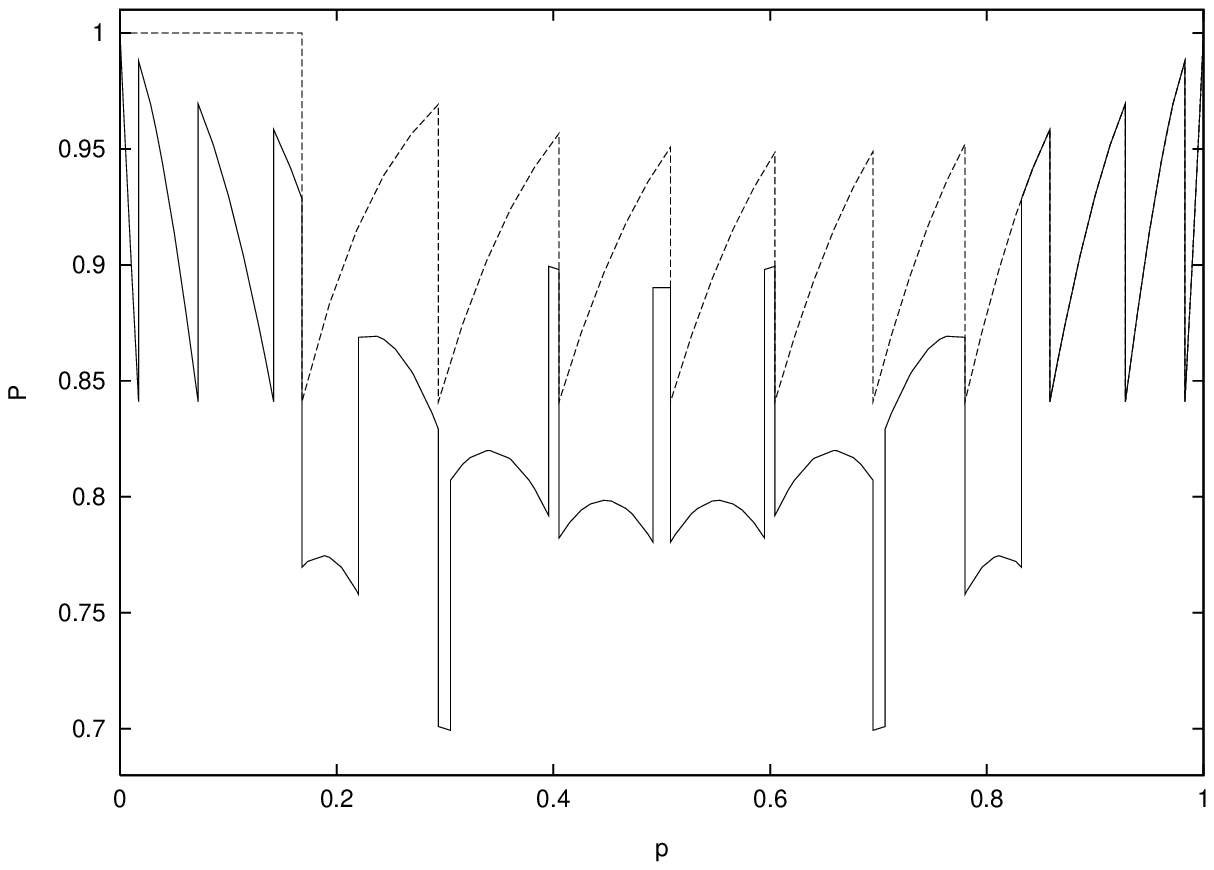}}
\caption{Confidence likelihoods for the Clopper-Pearson 
bounds~(\ref{CPbounds}). 
The parameters $N=10$ and $p^c_-=p^c_+=0.159$ are used. Upper, 
broken line: Confidence likelihood $P=P^c_+(p)$~(\ref{CPbounds}b) 
versus the true signal probability $p$.  Lower, full line: Confidence 
likelihood $P=P^c(p)$~(\ref{CPbounds}c) versus $p$.}
\end{figure}

The precise meaning of the bounds~(\ref{CPbounds})
is as follows: $p_-$ is the largest number such that (for
every feasible $p$) the probability for $p<p_-$ is less than $p^c_-$. 
Correspondingly, $p_+$ is the smallest number such that the probability
for $p>p_+$ is less than $p^c_+$. The other way round, 
$$ p\ge p_-\ {\rm with\ likelihood}\ P^c_- (p)\ge (1-p^c_-)
   \eqno{(\ref{CPbounds}a)} $$
and
$$ p\le p_+\ {\rm with\ likelihood}\ P^c_+ (p)\ge (1-p^c_+)\, .
   \eqno{(\ref{CPbounds}b)} $$
Therefore, for $p_-<p_+$ we find
$$ p\in [p_-,p_+]\ {\rm with\ likelihood}\ P^c (p) = P^c_- + P^c_+ - 1
   \ge (1-p^c_--p^c_+)\, .  \eqno{(\ref{CPbounds}c)} $$
It is instructive to illustrate these equations for a small value of 
$N$. Choosing $N=10$ and $p^c_-=p^c_+=0.159$, the precise $p$-dependence
of $P^c_+$~(\ref{CPbounds}b) and of $P^c$~(\ref{CPbounds}c) is depicted 
in figure~\ref{CPprob}. The equality $P^c_+(p)=1-p^c_+$ is assumed at
the discrete values $p=p_+(N_s)$, $N_s=0,1,...,N$. For example, as long
as $p\le p_+(0)$ holds, $p$ certainty will be smaller than any $p_+$
bound. As $p$ passes through $p_+(0)$, the probability $P_+(p)$ jumps
down to the value $1-p^c_+=0.841$. Subsequently $P^c_+(p)$ rises with
$p$ in the range $p_+(0)<p<p_+(1)$ until, at $p=p_+(1)$, the next jump
occurs, and so on. The corresponding graph for $P^c_- (p)$ follows from
$P^c_+(p)$ by reflection on the $p=0.5$ axis. The lower, full curve of
figure~\ref{CPprob} is obtained by combining both according to
eqn.(\ref{CPbounds}c).

We are interested in the more involved situation where signal and 
background can no longer be distinguished unambiguously. Instead, a
neural network or similar device yields statistical information by
tagging signals with efficiency $F_s$ and background data with 
efficiency $F_b$~~ ($0\le F_b\le 1$, $0\le F_s\le 1$ and, typically, 
$F_b\ll F_s$). Applying the network to all $N$ data results in $N^Y$,
($0\le N^Y\le N$) tagged data, composed of $N^Y=N^Y_s+N^Y_b$, where
$N^Y_s$ are the tagged signals and $N^Y_b$ are the tagged background 
data. Of course, the values for $N^Y_s$ and $N^Y_b$ are not known. 
Our task is to determine
confidence levels for the signal likelihood $p$ from the sole knowledge
of $N^Y$. We proceed by writing down the probability density of $N^Y$ 
for given $p$ and, subsequently, generalizing the Clopper-Pearson method.

First, assume fixed $N_s$. The probability densities of $N^Y_s$ and 
$N^Y_b$ are binomial and thus the probability density for $N^Y$
is given by the convolution
\begin{equation} \label{PYS}
P(N^Y|N_s) = \sum_{N^Y_s+N^Y_b=N^Y} b(N^Y_s|N_s,F_s)\, b(N^Y_b|N_b,F_b),\
N_b=N-N_s ~.  
\end{equation}
Summing over $N_s$ removes the constraint and the $N^Y$-probability
density, with $N$, $p$ fixed, is
\begin{equation} \label{PYp}
P(N^Y|N,p) = \sum_{N_s=0}^N b(N_s|N,p)\, P(N^Y|N_s)~.
\end{equation}
For given $p^c_-$, $p^c_+$ and $N^Y$, we define confidence limits $p_-$ 
and $p_+$ in analogy with equation~(\ref{CPbounds})
\begin{equation} \label{BRbounds}
  p^c_-=\sum_{k=N^Y}^N P(k|N,p_-)\ ~~{\rm and}\ 
~~p^c_+=\sum_{k=0}^{N^Y} P(k|N,p_+) ~.
\end{equation}
Their meaning is as already outlined by 
equations~(\ref{CPbounds}a-\ref{CPbounds}c). Choosing $F_s=0.9$, $F_b=0.2$ and
the other parameters as before, figure~3 illustrates these equations 
for the new situation. The interpretation is as for figure~\ref{CPprob}
with two remarkable exceptions:

\begin{figure}[b] \label{BRprob}
\centerline{\psfig{figure=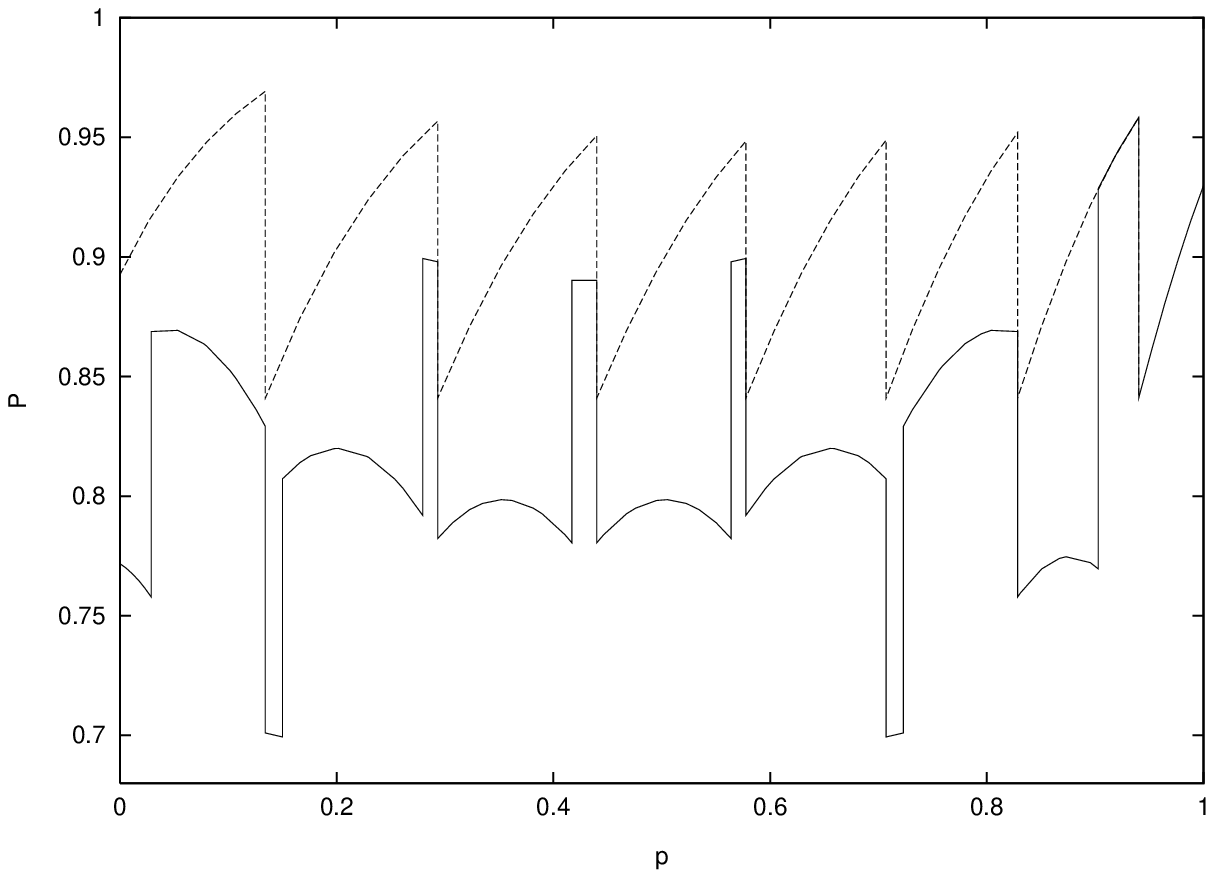}}
\caption{Confidence likelihoods for the generalized Clopper-Pearson
bounds~(\ref{BRbounds}). The parameters $F_s=0.9$, $F_b=0.2$ and those of 
figure~\ref{CPprob} are used. Upper, broken line: Confidence 
likelihood $P=P^c_+(p)$~(\ref{CPbounds}b) 
versus the true signal probability $p$.  Lower, full line: Confidence 
likelihood $P=P^c(p)$~(\ref{BRbounds}c) versus $p$.}
\end{figure}

\begin{description}

\item{(i)} It may happen that eqn.(\ref{BRbounds}) has no solutions 
$p_-(N^Y)$ for certain $N^Y=N,N-1,...$ or no solutions $p_+(N^Y)$ for
certain $N^Y=0,1,...$~. The reason is that, due to the NN, the result 
is sufficiently unlikely for all $p$. One may then either decrease
$p^c_-$ or $p^c_+$ or discard the entire analysis. The parameter values
of figure~3 are chosen such that there is no solution 
$p_+(0)$. Consequently, there is no longer a range of small $p$-values
with $P^c_+(p)=1$. Such exotic NN output (here $N^Y=0$) is by definition
rare.

\item{(ii)} For $F_s+F_b\ne 1$, the function $P^c_-(p)$ is no longer a 
reflection of $P^c_+(p)$. This is shown in figure~3, where $P^c (p)$ turns
out to be no longer symmetric. In fact, on the r.h.s of figure~3 we observe 
the same feature as in figure~2: The upper and lower curves agree due to
$P_-(p)=1$ in this range.

\end{description}

In summary, the bounds $[p_-,p_+]$ obtained with $p^c_-=p^c_+=0.159$
guarantee the standard one error bar confidence probability of 68.2\% 
for every single $p$-value and for almost all $p$ the actual confidence 
will be better. However, the one-sided bounds cannot be improved without
violating the requested confidence probability for some $p$-values.
In the same way, bounds calculated with $p^c_-=p^c_+=0.023$ ensure the
standard two error bar confidence level of 95.4\% or better, and so on.
It should be noted that, for $p\ne 0$ and $p\ne 1$, the deviations from 
the requested confidence probabilities tend to decrease in the limit of 
large statistics.
\hfill\break

\section{Large Data Sets with Few Signals}

We now assume the values of figure~1 to demonstrate the approach in a 
limit which is of particular interest for experimental high energy physics 
applications. With $N=26k$ and $N_s=130$ one gets the Clopper-Pearson
confidence limits
$$ 0.00456 \le p \le 0.00547~ \mbox{for}~ p^c_-=p^c_+=0.159\, , $$
$$ 0.00416 \le p \le 0.00595~ \mbox{for}~ p^c_-=p^c_+=0.023\, . $$
Next, we assume that the only information about the signals is
provided by some NN output, where we use two sets of efficiencies,
inspired by~\cite{Be96}. 

First, we consider $F_s=0.5$ and $F_b=0.005$.
Figure~4 depicts the tag probability density
$P(N^Y|N_s)$, see~(\ref{PYS}), for three different values of $N_s$:
$0,\ 130$ and $260$. There is almost no overlap and, consequently,
we expect that clear identification of a positive signal can be
achieved. For $N_s=130$ the central $N^Y$ values are located around
$N_sF_s+N_bF_b=130\, F_s + (26000-130)\, F_b = 194.35$. Using 
$N^Y=194$, iteration of equation~(\ref{BRbounds}) yields the confidence limits
$$ 0.00389 \le p \le 0.00613~ \mbox{for}~ p^c_-=p^c_+=0.159\, , $$
$$ 0.00289 \le p \le 0.00729~ \mbox{for}~ p^c_-=p^c_+=0.023\, . $$
The computational demand for these results was less than two hours of CPU 
time on a DEC 3000 Alpha 600 workstation, where it is important to store
frequently used coefficients in RAM.

\begin{figure}[b]
\centerline{\psfig{figure=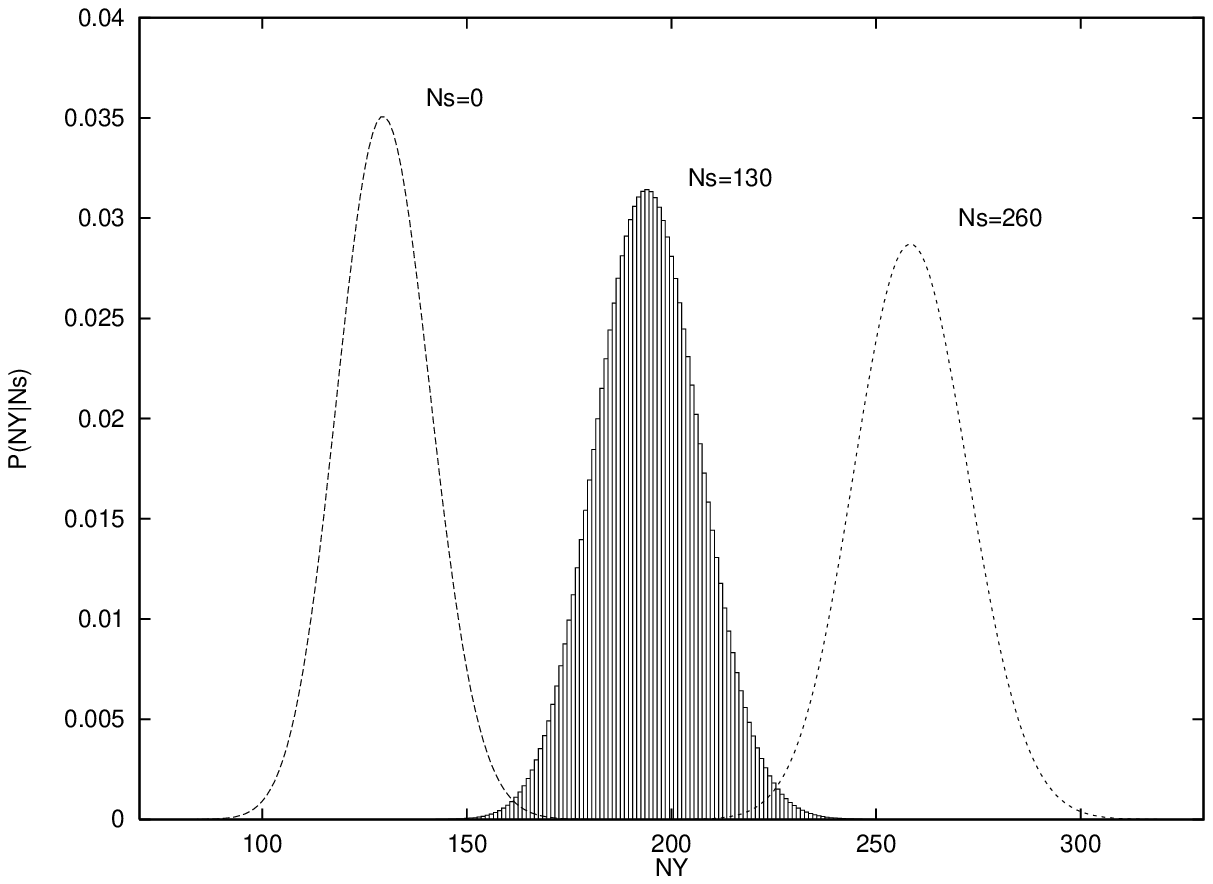}}
\caption{Probability densities to get $N^Y$ data from the NN employing
the efficiencies $F_b=0.005, F_s=0.5$ and assuming $N_s$ signals in the
original $26k$ data set.} 
\end{figure}

Let us reduce the signal efficiency to $F_s=0.1$ and keep the background 
(in)efficiency unchanged. Figure~5 depicts the new tag probability 
densities. We find considerable overlap and expect that $p=0$ can no 
longer be excluded. The central $N^Y$ values are now located around 
$142.35$. Using $N^Y=142$, iteration of equation~(\ref{BRbounds}) gives
$$ 0.00005 \le p \le 0.010~~ \mbox{for}~ p^c_-=p^c_+=0.159\, , $$
$$ 0.00000 \le p \le 0.0153~ \mbox{for}~ p^c_-=p^c_+=0.023\, . $$
The latter case should be supplemented by the explicit probability 
for $p=0$, estimated in the next section.
\hfill\break

\begin{figure}[t]
\centerline{\psfig{figure=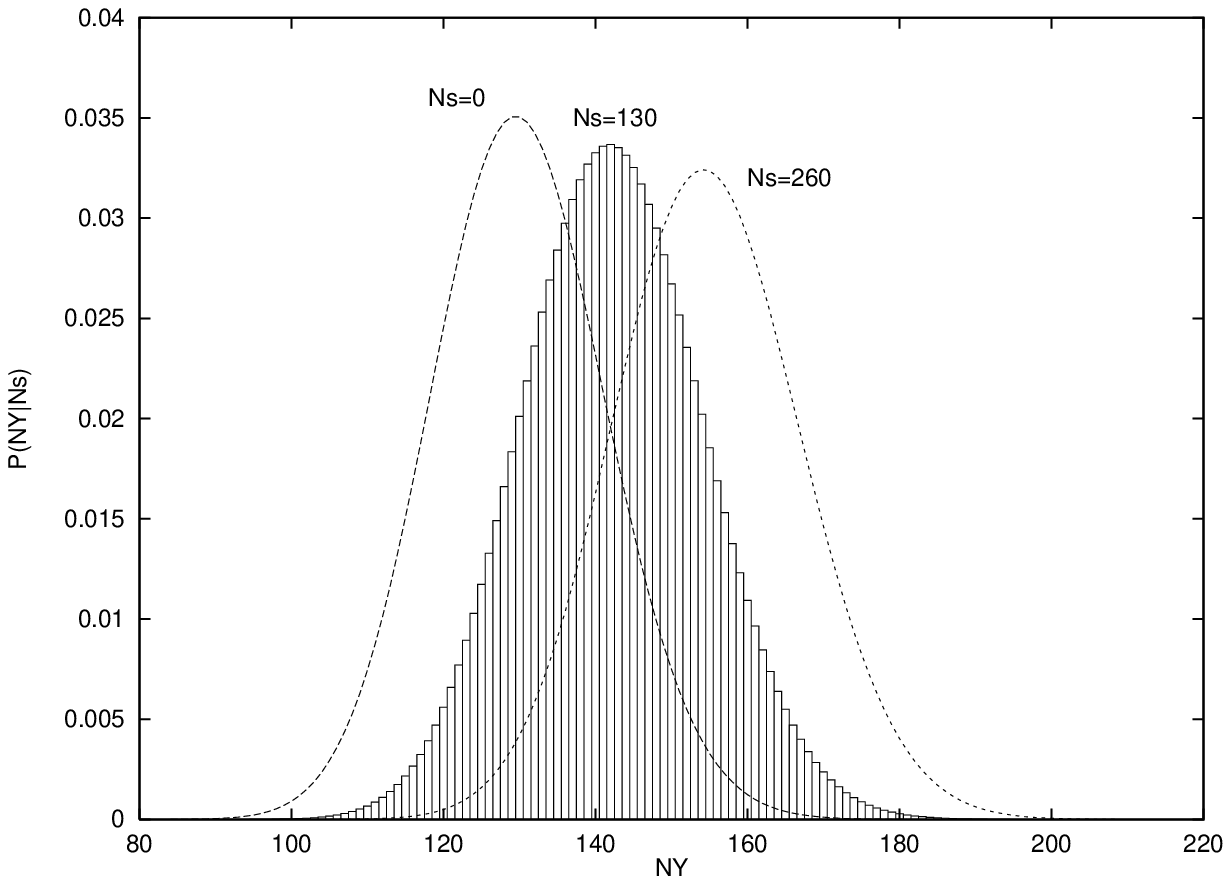}}
\caption{As figure~4, but with signal network efficiency $F_s=0.1$.}
\end{figure}

\section{Signal Probability Distributions}

Equation~(\ref{BRbounds}), or of course~(\ref{CPbounds}) when applicable, 
can be used to sandwich the {\it a-posteriori} signal probability
distribution $F(p)=F(p|N,N^Y)$ between lower and upper bounds. Namely, 
it is easy to see that
\begin{equation} \label{Fpbounds}
F_1(p) = 1-p^c_+(p) = \sum_{k=N^Y+1}^N P(k|N,p) \ \le\ 
F(p)\ \le\ F_2(p) = p^c_-(p)   = \sum_{k=N^Y}^N P(k|N,p)\, .
\end{equation} 
For their numerical evaluation the sums should be re-written as
$F_2=1-\sum_{k=0}^{N^Y-1} P(k|N,p)$ and $F_1=F_2-P(N^Y|N,p)$.
Figure~6 depicts these functions for the previously discussed
examples $N^Y=142$ and $194$. Upper and lower bounds are seen
close together, such that $F(p)=(F_1(p)+F_2(p))/2$ would be a 
reasonable working approximation. The corresponding probability 
densities are the derivatives with respect to $p$. Their numerical 
calculation is straightforward when analytical expressions for the 
derivatives of the binomial coefficients in equation~(\ref{PYp}) are
used. Figure~7 exhibits the results, $P_1(p|N,N^Y)$ and 
$P_2(p|N,N^Y)$. At $p=0$ the probability densities have 
$\delta$-function contributions
$$ P_i (p)\ =\ F_i(0)\, \delta (p) + ...\, ,\ (i=1,2)
~~{\rm with}~~ F_1(0)=0.136\ {\rm and}\ F_2(0)=0.156\, .$$

\begin{figure}[t]
\centerline{\psfig{figure=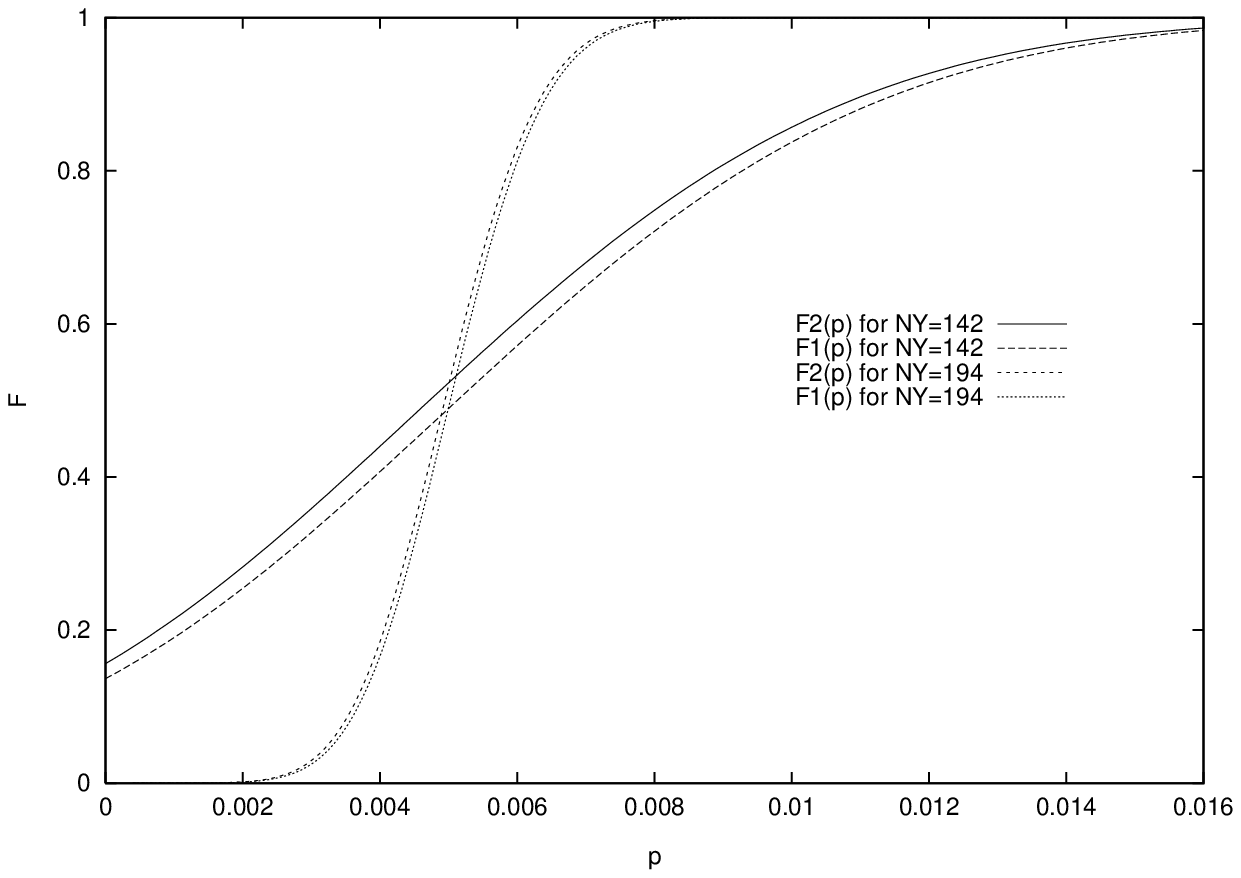}}
\caption{{\it A-posteriori} signal probability distributions
(upper and lower bounds).}
\end{figure}

\begin{figure}[t]
\centerline{\psfig{figure=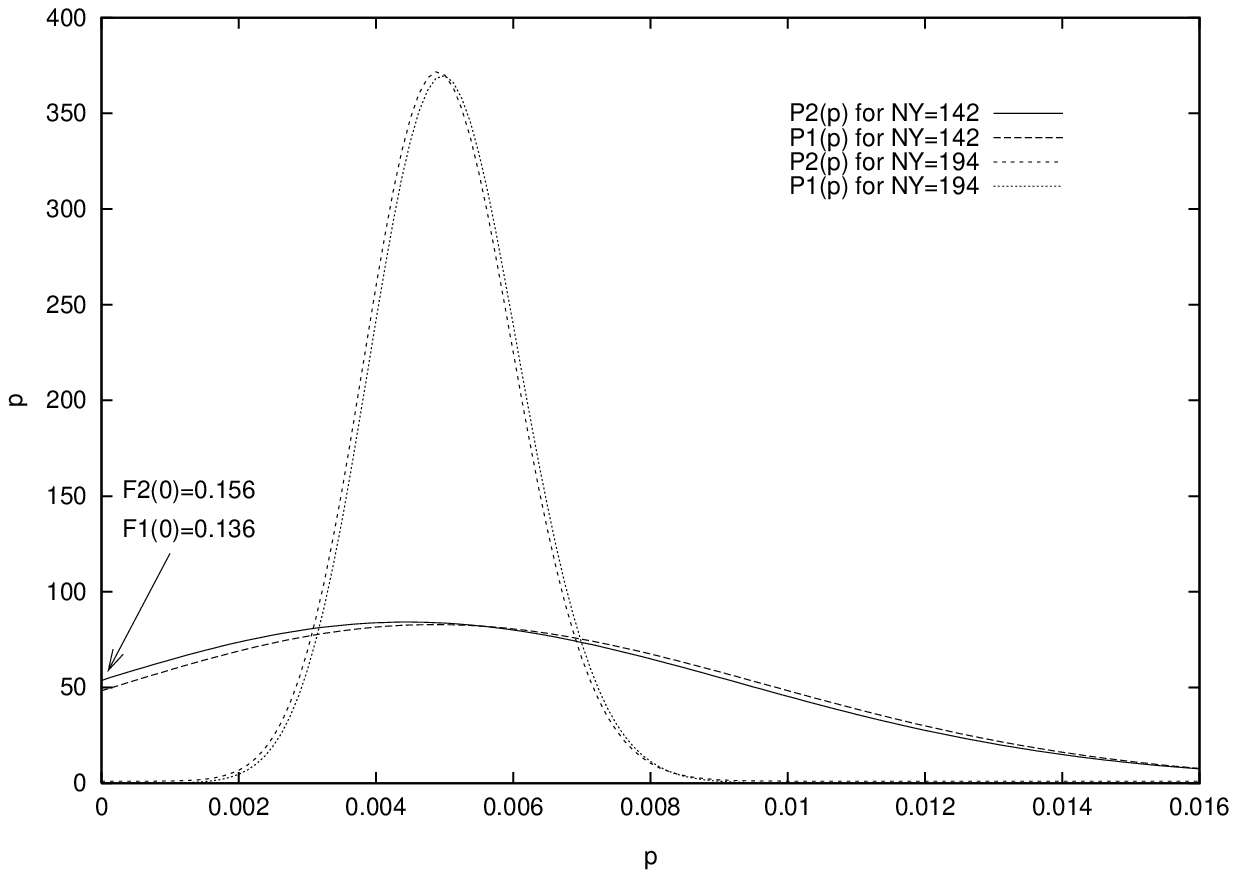}}
\caption{{\it A-posteriori} signal probability densities
(corresponding to upper and lower bounds of the distribution
functions in figure~6).}
\end{figure}

In addition, or alternatively to the outlined approach, one may be 
interested to find for $N_s=0,1,...,N$ the probabilities that there 
are $N_s$ signals in the data. That could be done using the 
probability densities $P_i(p|N,N^Y)$, $(i=1,2)$, but a calculation 
starting off from $P(N^Y|N_s)$, equation~(\ref{PYS}), is far more 
direct. In particular, it may sometimes be of advantage that $N_s$, 
in contrast to $p$, is a discrete variable.
Let us now denote the probability distribution for signals by
$F(N_s)=F(N_s|N^Y)$. Lower and upper bounds are
\begin{equation} \label{FNSbounds}
F_1(N_s) = \sum_{k=N^Y+1}^N P(k|N_s)\ \le\ 
F(N_s)\ \le\ F_2(N_s) = \sum_{k=N^Y}^N P(k|N_s)\, .
\end{equation} 
Despite of using the same symbols $F$, $F_1$ and $F_2$, the functions 
in equation~(\ref{Fpbounds}) and (\ref{FNSbounds}) are, of course, 
different. By definition, $F_2(N_s)$ is the likelihood that $N_s$
signals could have produced the observed $N^Y$ or a greater one.
Therefore, the likelihood that either of $k=0,1,...,N_s$ is correct
is less or equal the value $F_2(N_s)$, {\it i.e.} $F_2(N_s)$ is
an upper bound of the {\it a-posteriori} distribution function $F(N_s)$.
Similarly, $1-F_1(N_s)$ is an upper bound on the likelihood that
either of $k=N_s,N_s+1,...,N$ is correct. Consequently, $F_1(N_s)$ is
a lower bound of $F(N_s)$.  Figure~8 depicts theses bounds for our 
standard examples $N^Y=142$ and $194$. The similarity with figure~6
is no coincidence, as $p$ determines $N_s$ up to fluctuations of order
$1/\sqrt{N}$. Probability densities $P_i(N_s|N^Y)$ are defined by
$$F_i (N_s|N^Y)\ =\ \sum_{k=0}^{N_s} P_i(k|N^Y)\, ,~~(i=1,2)\, .$$
Defining $F_i(-1)=0$, they follow recursively
\begin{equation} \label{PNS}
P_i (N_s|N^Y)\ =\ F_i(N_s) - F_i(N_s-1),\ (N_s=0,1,...,N)\, .
\end{equation} 
Figure~9 exhibits the results.
Once the probability densities $P_i(N_s|N^Y)$, $\sum_{N_s=0}^N P_i(N_s|N^Y)=1$ 
are known, confidence limits can also be calculated from the subsequent 
generalization of the Clopper-Pearson~(\ref{CPbounds}) approach:
\begin{equation} \label{mppm}
p^c_-=\sum_{N_s=0}^N P_2(N_s|N^Y)\sum_{k=N_s}^{N}\, b(k|N,p_-) ,~~~ 
p^c_+=\sum_{N_s=0}^N P_1(N_s|N^Y)\sum_{k=0}^{N_s}\, b(k|N,p_+) ~.
\end{equation}
These equations involve nothing, but weighting the binomial 
Clopper-Pearson sums with the appropriate probabilities $P(N_s|N^Y)$.
They re-produce the bounds (\ref{BRbounds}) identically, as was numerically
checked for our examples of section~3.

\begin{figure}[t]
\centerline{\psfig{figure=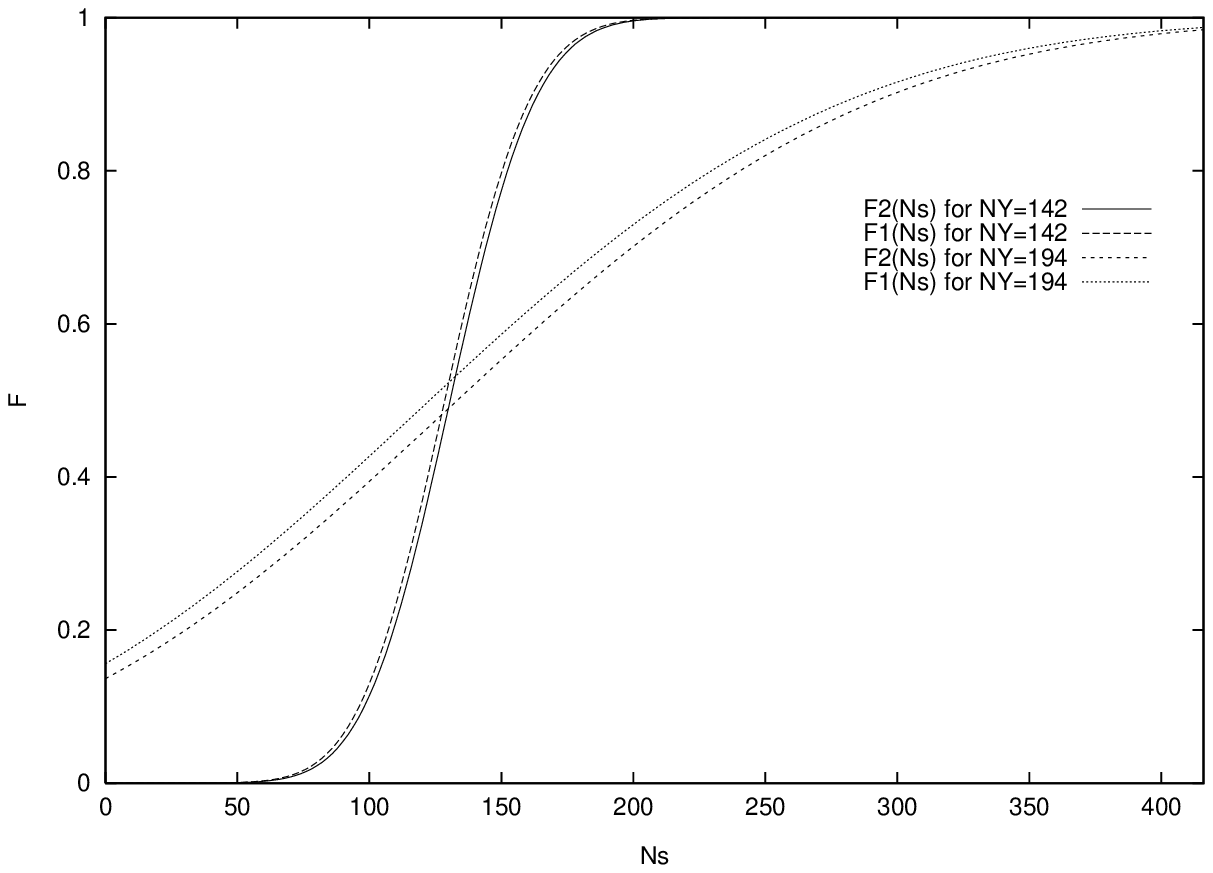}}
\caption{{\it A-posteriori} distributions for the number of signals
(upper and lower bounds).}
\end{figure}

\begin{figure}[t]
\centerline{\psfig{figure=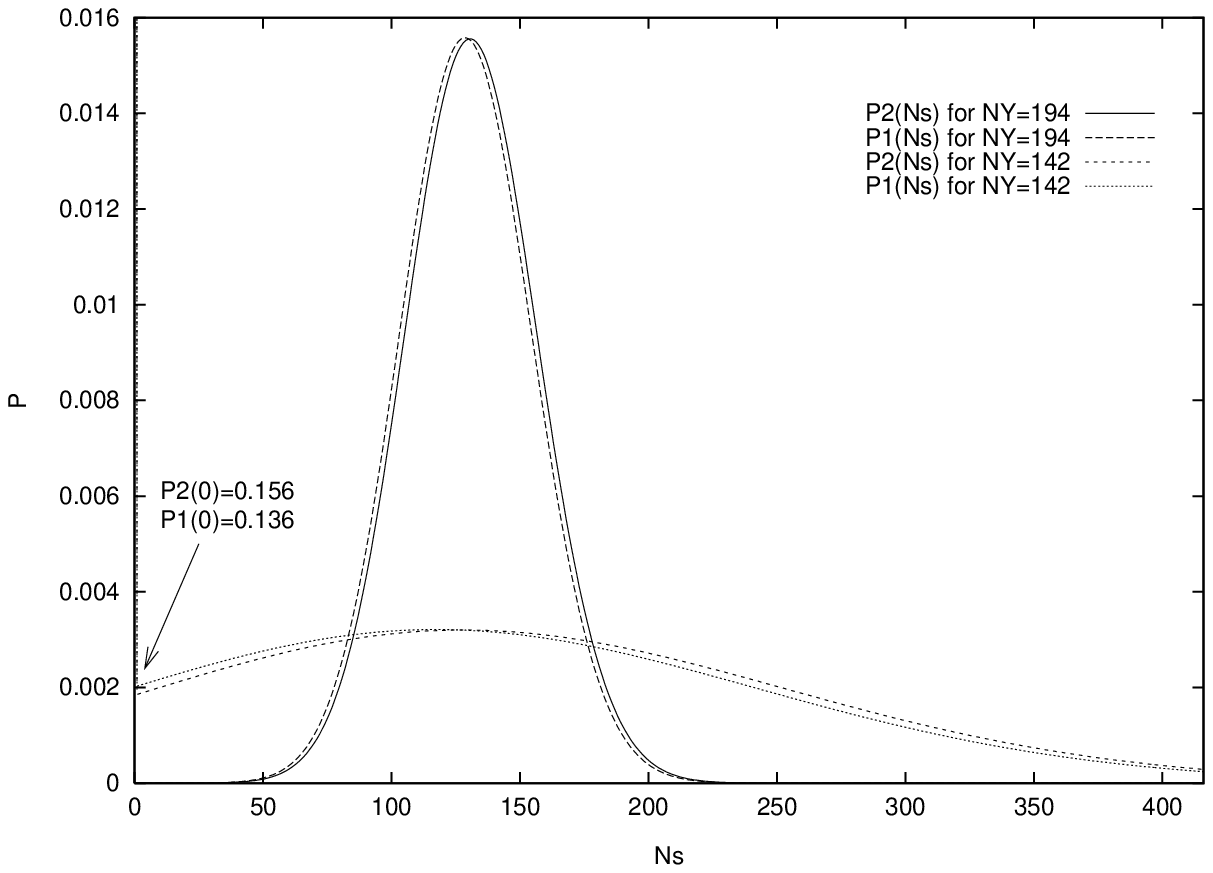}}
\caption{{\it A-posteriori} probability densities for the number
of signals (corresponding to upper and lower bounds of the distribution
functions in figure~8).}
\end{figure}
\hfill\break

\section{Bayesian Approach}

Our construction invokes the {\it a-priori} known fact that the number of 
signals is in the range $0\le N_s \le N$. It is popular (for reviews
see \cite{Lor90,Bay95}), and sometimes quite successful, to  make additional 
assumptions in form of {\it a-priori} likelihoods. This can be motivated
by a look at figures~2 and~3. For almost all $p$ the confidence is better
than the desired 68.2\%. If an {\it a-priori} likelihood is known,
the Bayesian approach yields a confidence of precisely 68.2\%. The debate 
is about using {\it a-priori} likelihoods in situations where they are
{\it not} known. Reasonable guesses can apparently be made in many
situations. However, false results are obtained when such a guess is
in contradiction with the data, which may not always be trivial to 
uncover. An example is given here.

In our situation, one would be tempted to impose an {\it a-priori} 
likelihood on either $p$ or $N_s$. For instance, invoking the maximum-entropy 
principle \cite{Jay82} leads to constant {\it a-priori} probability densities
\begin{equation}  \label{P0Ns}
\rho^0 (p) = 1 ~~{\rm or}~~
 P^0(N_s) = {1\over N+1} ~~{\rm for}~~ 0\le N_s\le N\, .
\end{equation}
For simplicity we focus on the latter case. (Using the result from 
\cite{Bay95a} it would also be straightforward to work out the other one.)
As before, $N^Y$ is determined by measurements and NN analysis. Under the 
assumption
(\ref{P0Ns}) for $P^0(N_s)$, the Bayesian theorem implies the 
{\it a-posteriori} probability
\begin{equation}  \label{PbNs}
 P(N_s|N^Y)= const.\ P(N^Y|N_s)\, ,
\end{equation}
with $P(N^Y|N_s)$ given by equation~(\ref{PYS}) and the constant follows 
from the normalization $\sum_{N_s=0}^N P(N_s|N^Y) = 1$. In case of
$N^Y=194$ the result agrees very well with that depicted in figure~9.
However, this is not true for $N^Y=142$, see figure~10, where 
the probability density of figure~9 is compared with the Bayesian result.

\begin{figure}[t]
\centerline{\psfig{figure=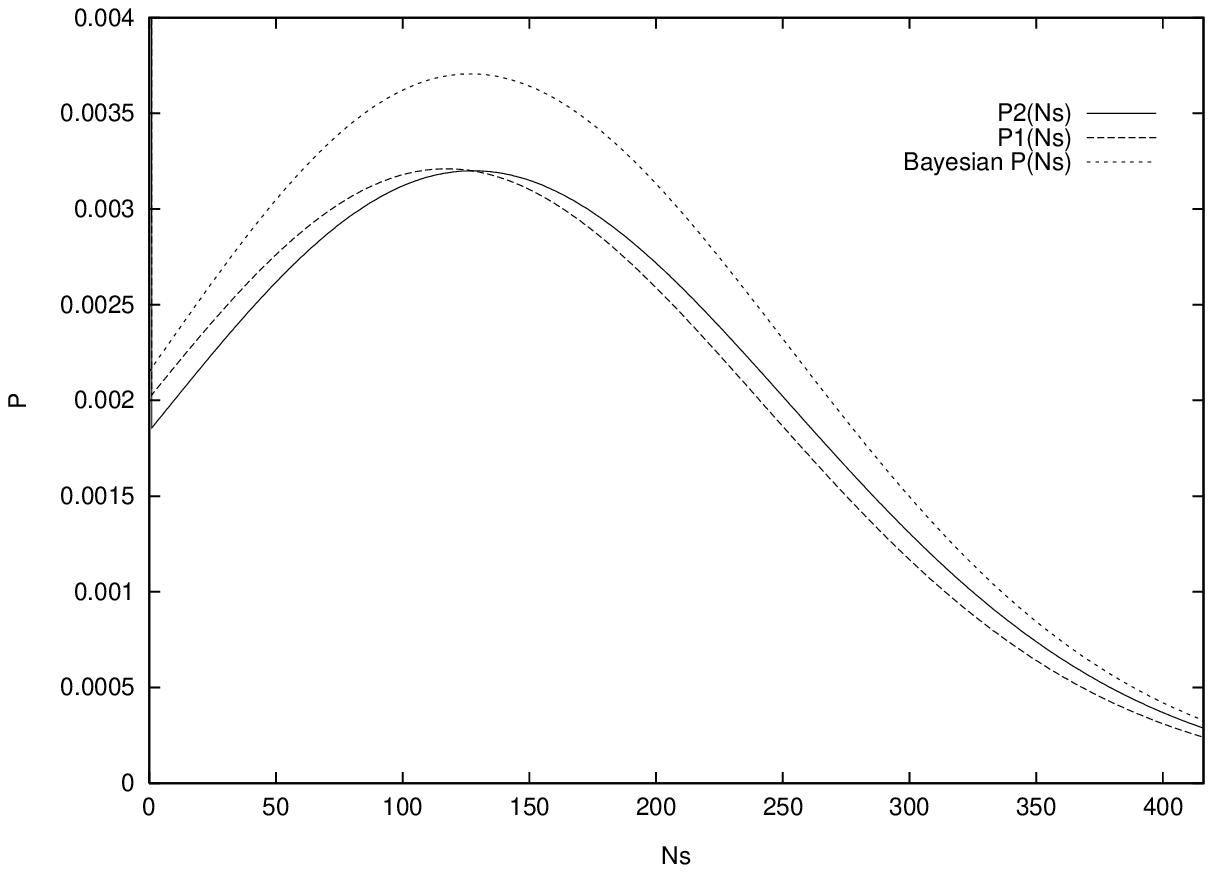}}
\caption{$N^Y=142$: Comparison of the Bayesian (maximum entropy)
{\it a-posteriori} probability density $P(N_s)$
with $P_1(N_s)$ and $P_2(N_s)$ of figure~9.}
\end{figure}

Whereas for strong signal identification the difference with our approach 
is practically negligible, it is significant for weak (or no) signal 
identification. The Bayesian probability for $p=0$ is only 0.00215,
implying that the Bayesian distribution function violates the rigorous 
$F_1(N_s)$ bound. The reason is obvious: It is not clear what {\it a-priori} 
probability one should assign to the situation that there is no effect 
at all. That is why $N_s=0$ does not compete on the same level as 
the numbers $N_s\ge 1$. For one of the the situations, we have in mind, 
no effect at all would mean that there is no top quark. One could assign 
a finite
{\it a-priori} probability to this possibility, but whether this is
10\%, 50\% or 90\% would be highly arbitrary. Actually it does not even 
work: As the top quark has already be found \cite{D0}, one may argue in 
favor of the given {\it a-priori} likelihood with the argument that 
the $N_s=0$ is certainly very small. However, this leads to an 
overestimation of large signal probabilities, as the {\it a-posteriori} 
$N_s=0$ likelihood becomes incorrectly re-distributed.
\hfill\break

\section{Conclusions}
We have calculated confidence limits of an unknown signal likelihood for 
the situation where few signals occur in a large number of events. The only
input used were neural network efficiencies for tagging signal and background
events as well as the number of data the network selects. The extension of
our approach from the binomial to the multinomial case, {\em i.e.}
to more than two different types of data (signal and background) is
certainly possible.

In typical applications the efficiencies $F_s$ and $F_b$ may not be
known exactly either. Instead, a number of training sets $(j=1,...,J)$
may exist, each giving somewhat different efficiencies $F^j_s,\, F_b^j$.
We think that in this situation a bootstrap type of approach~\cite{Ef87}
can be applied and that the probability density~(\ref{PNS}) provides
a suitable starting point. We can linearly combine different probability
densities to an ultimate one
$$ P_i (N_s) = J^{-1} \sum_{j=1}^J P^j_i (N_s|N^Y_j)\, , (i=1,2)\, ,$$
and proceed with $P_i(N_s)$ as discussed in section~4.

Finally, to involve conjectured {\it a-priori} likelihoods may in many 
situations be unavoidable and, actually, be quite successful. In our case:
When a clear, positive signal identification is possible, we find practically 
no difference between a Bayesian maximum entropy and our approach. However, 
our example of weak signal identification shows that {\it a-priori} 
likelihoods are better avoided when a rigorous alternative exists.
\hfill\break

\end{document}